\documentclass{article}
\usepackage{latexsym}

\def\dag{\dagger} \def\pd{\partial} \def\pp{\prime} \def\a{\alpha} \def\b{\beta} \def\dl{\delta} \def\s{\sigma}  \def\eps{\epsilon}  \def\lam{\lambda}  \def\gm{\gamma}  \def\om{\omega}  \def\sq{\sqrt} \def\fr{\frac} \def\half{\frac{1}{2}}
\def\rb{{\rm b}} \def\rc{{\rm c}} \def\gh{{\rm gh}}  \def\BRST{{\rm BRST}} 
 
\def\hg{{\hat g}}    \def\hnb{{\hat \nabla}}  \def\hDelta{{\hat \Delta}}  \def\hR{{\hat R}}      

\def\calA{{\cal A}} \def\calB{{\cal B}} \def\calR{{\cal R}} \def\calP{{\cal P}} \def\calV{{\cal V}} \def\calO{{\cal O}}
\def\bx{{\bf x}} \def\by{{\bf y}} \def\bk{{\bf k}}
 \def\P{{\rm P}} 
  
  \def\P{{\sf P}}   

\def\lap3{~| \!\!\! \partial^2} \def\dlap3{~| \!\!\! \partial^4} \def\invlap3{~| \!\!\! \partial^{-2}}
\def\l:{: \!}\def\r:{\! :}
\def\lang{\langle} \def\rang{\rangle}
\def\d3x{d^3 {\bf x}~}

\begin{document}

\begin{titlepage}

\begin{flushright}
March 2012
\end{flushright}

\vspace{5mm}

\begin{center}
{\Large {\bf BRST Invariant Higher Derivative Operators in 4D Quantum Gravity based on CFT}} 
\end{center}

\vspace{5mm}

\begin{center}
{\sc Ken-ji Hamada}\footnote{E-mail address: hamada@post.kek.jp; URL: http://research.kek.jp/people/hamada/}
\end{center}

\begin{center}
{\it Institute of Particle and Nuclear Studies, KEK, Tsukuba 305-0801, Japan} \\ and \\
{\it Department of Particle and Nuclear Physics, The Graduate University for Advanced Studies (Sokendai), Tsukuba 305-0801, Japan}
\end{center}

\begin{abstract}
We continue the study of physical fields for the background free 4D quantum gravity based on the Riegert-Wess-Zumino action, developed in Phys. Rev. D {\bf 85} (2012) 024028 \cite{hamada12a}. The background free model is formulated in terms of a certain conformal field theory on $M^4$ in which conformal symmetry arises as gauge symmetry, namely diffeomorphism invariance. In this paper, we construct the physical field operator corresponding to any integer power of Ricci scalar curvature in the context of the BRST quantization. We also discuss how to define the correlation function and its physical meanings.
\end{abstract}

\end{titlepage}

\section{Introduction}
\setcounter{equation}{0}
\noindent

Conformal field theory (CFT) \cite{isz, bz, kluwer} is generally known to be a scale invariant field theory realized at fixed points of renormalization group flows \cite{wk}. Quantum gravity is also described as a certain quantum field theory with exact conformal invariance in another sense, because the metric field has the conformal factor and thus conformal invariance is realized as a part of diffeomorphism invariance when this factor fully fluctuates quantum mechanically. Really, it is well-known that 2D quantum gravity is described in terms of the CFT called the Liouville theory \cite{polyakov, kpz, dk, seiberg, gl}.

The four dimensional quantum gravity we will study here is described in terms of such a CFT in the ultraviolet (UV) limit, which is defined by the perturbation theory about a conformally flat configuration characterized by the expansion of the metric field \cite{hamada02, hamada09a}
\begin{equation}
     g_{\mu\nu}= e^{2\phi} \left( \hg_{\mu\nu} + t h_{\mu\nu} + \cdots \right) ,
       \label{metric decomposition}
\end{equation}
where $tr(h)=\hg^{\mu\nu}h_{\mu\nu}=0$ and $\hg_{\mu\nu}$ is the background metric, which is practically chosen to be the Minkowski metric $\eta_{\mu\nu}=(-1,1,1,1)$. The coupling constant $t$ is dimensionless, which indicates the asymptotic freedom. The conformal factor is treated exactly without introducing its own coupling constant. The derived CFT is described by the combined system of the Riegert-Wess-Zumino action \cite{riegert, am, amm92, hs, hamada12a, hamada12b} and the Weyl action. The conformal symmetry is then realized as the residual diffeomorphism symmetry left after the gauge fixing such that gauge degrees of freedom reduce to the 15 conformal Killing vectors $\zeta^\mu$ satisfying $\pd_\mu \zeta_\nu + \pd_\nu \zeta_\mu - \eta_{\mu\nu} \pd^\lam \zeta_\lam/2 =0$.

In this paper, we continue the study of physical field operators developed in \cite{hamada12a} in the context of the Becchi-Rouet-Stora-Tyupin (BRST) quantization \cite{brs, kugo, kato, fms}. The BRST transformation we study here is obtained by replacing $\zeta^\mu$ in the conformal transformations with the corresponding gauge ghost $c^\mu$ as 
\begin{eqnarray}
     \dl_{\rm B} \phi &=& c^\lam \pd_\lam \phi + \fr{1}{4} \pd_\lam c^\lam ,
             \nonumber \\ 
    \dl_{\rm B} h_{\mu\nu} &=& c^\lam \pd_\lam h_{\mu\nu} 
              + \half h_{\mu\lam} \left( \pd_\nu c^\lam - \pd^\lam c_\nu \right)
              + \half h_{\nu\lam} \left( \pd_\mu c^\lam - \pd^\lam c_\mu \right) ,
             \nonumber \\
    \dl_{\rm B} c^\mu &=& c^\lam \pd_\lam c^\mu ,  
           \label{BRST transformation}       
\end{eqnarray}
where the BRST transformation of the gauge ghost is added. The generators of transformations $\dl_{\rm B} \phi$ and $\dl_{\rm B} h_{\mu\nu}$ are derived from the Riegert-Wess-Zumino action and the Weyl action, respectively. Unlike usual CFT, this conformal invariance is imposed on the field as well as the vacuum because it is a gauge symmetry. Thus, it gives stringent constraints to physical quantities.

This paper is presented as follows. In the next section, we summarize the generators of diffeomorphism symmetry we use here. In Section 3, we reinvestigate the physical operator corresponding to the cosmological constant term given in the previous paper \cite{hamada12a} and then rewrite it in the locally BRST invariant form. The quantum Ricci scalar operator is constructed in Section 4. In Section 5, the argument is generalized to higher derivative operators. We here construct the physical operator corresponding to the integer power of Ricci scale curvature. In Section 6, we give remarks on correlation functions among physical fields. Section 7 is devoted to conclusion.

In the following, the spacetime coordinate is described by $x^\mu=(\eta, x^i)$ and $x^2=x_\mu x^\mu =-\eta^2 + \bx^2$. The d'Alembertian is denoted by $\pd^2= \pd^\mu \pd_\mu=-\pd_\eta^2 + \lap3$, where $\lap3=\pd_i \pd^i$ is the Laplacian of three dimensional space.

\section{BRST Operator Imposing Diffeomorphism Invariance}
\setcounter{equation}{0}
\noindent

Nonperturbative dynamics of the Riegert field $\phi$ is governed by the Riegert-Wess-Zumino action defined by \cite{riegert}
\begin{eqnarray}
 S_{\rm RWZ}= -\fr{b_1}{(4\pi)^2} \int d^4 x \sq{-\hg} \left\{ 2\phi \hDelta_4 \phi 
                            + \left( \hat{G}_4 - \fr{2}{3}\hnb^2 \hR \right) \phi \right\} , 
                \label{Riegert action}
\end{eqnarray}
where $G_4$ is the Euler density and $\sq{-g}\Delta_4$ is the conformally invariant fourth-order differential operator. The quantities with the hat on them are defined in terms of the background metric $\hg_{\mu\nu}$. Here, we use the Minkowski background $M^4$ and thus the conformally invariant differential operator on the background is given by $\pd^4$. This action is induced from the measure as in the case of Liouville action in 2D quantum gravity. The coefficient $b_1$ has the physically correct sign of positive. In this paper, we consider the Riegert sector only. For the Weyl sector, see the previous paper \cite{hamada12a}.

The Riegert field is quantized \cite{am, amm92, hs} following the Dirac's procedure by introducing new variable $\chi=\pd_\eta \phi$ \cite{hamada12a}. The four canonical variables are then given by $\chi$ and $\phi$ and their conjugate momenta
\begin{eqnarray}
   \P_\chi = - \fr{b_1}{4\pi^2} \pd_\eta \chi  , \qquad
   \P_\phi = - \pd_\eta \P_\chi - \fr{b_1}{2\pi^2} \lap3 \chi . 
         \label{conjugate momentum}
\end{eqnarray}
The canonical commutation relations are set as $[ \phi(\eta, \bx), \P_\phi(\eta, \bx^\pp) ] = [ \chi(\eta, \bx), \P_\chi(\eta, \bx^\pp) ] = i \dl_3 (\bx -\bx^\pp)$ and otherwise vanishes. In Appendix A, we briefly summarize various things about the quantized Riegert field.

We here summarize the generator of conformal symmetry realized as diffeomorphism invariance at the UV limit and the BRST operator imposing this invariance, derived in \cite{hamada12a}. First, the generators of Poincar\'{e} algebra are given by 
\begin{eqnarray}
     P_0 &=& \int \d3x \calA, \qquad  P_j = \int \d3x \calB_j ,
            \nonumber \\
     M_{0j} &=& \int \d3x \left[ -\eta \calB_j - x_j \calA  - \l: \P_\chi \pd_j \phi \r: \right], 
              \nonumber \\
     M_{ij} &=& \int \d3x \left[  x_i \calB_j -x_j \calB_i \right] ,
            \label{Poincare generator}
\end{eqnarray}
where $\l:~ \r:$ denotes the normal ordering and the local operators $\calA$ and $\calB_j$ are defined by
\begin{eqnarray}
    \calA &=& -\fr{2\pi^2}{b_1} \l: \P_\chi^2 \r: + \l: \P_\phi \chi \r:
              + \fr{b_1}{8\pi^2} \left[ 2 \l: \chi \lap3 \chi \r: + \l: \lap3 \phi \lap3 \phi \r: \right] ,
             \nonumber \\
    \calB_j &=& \l: \P_\chi \pd_j \chi \r: + \l: \P_\phi \pd_j \phi \r: . 
        \label{local operators A and B_j}          
\end{eqnarray}
Here, $P_0=H$ is the Hamiltonian. The generators of dilatations and special conformal transformations are given by
\begin{equation}
     D = \int \d3x  \left[ \eta \calA + x^k \calB_k  + \l: \P_\chi \chi \r: + \P_\phi \right] 
         \label{dilatation generator}
\end{equation}
and
\begin{eqnarray}
    K_0 &=& \int \d3x \biggl\{ 
               \left( \eta^2 + \bx^2 \right) \calA + 2\eta x^k \calB_k 
               + 2\eta \l: \P_\chi \chi \r: 
               + 2 x^k \l: \P_\chi \pd_k \phi \r: 
                 \nonumber \\
        && \qquad\qquad 
               - \fr{b_1}{4\pi^2} \left[ 2 \l: \chi^2 \r: + \l: \pd_k \phi \pd^k \phi \r: \right]
              + 2\eta \P_\phi + 2 \P_\chi  \biggr\} ,
                \nonumber \\
   K_j &=& \int \d3x  \biggl\{
               \left( -\eta^2 + \bx^2 \right) \calB_j  -2 x_j x^k \calB_k
               - 2\eta x_j \calA   - 2x_j \l: \P_\chi \chi \r:  
                  \nonumber \\
       && \qquad\qquad 
               - 2\eta \l: \P_\chi \pd_j \phi \r:   -\fr{b_1}{2\pi^2} \l: \chi \pd_j \phi \r: 
               - 2x_j \P_\phi 
             \biggr\} .
           \label{special conformal transformation generator}
\end{eqnarray}
The linear terms in $D$ and $K_\mu$ generate the shift term in the gauge transformation of Riegert field (\ref{BRST transformation}). These 15 generators form the closed algebra of conformal symmetry,
\begin{eqnarray}
    \left[ P_\mu, P_\nu \right] &=& 0 ,  \qquad
    \left[ M_{\mu\nu}, P_\lam \right] = -i \left( \eta_{\mu\lam} P_\nu - \eta_{\nu\lam} P_\mu \right),
          \nonumber \\
    \left[ M_{\mu\nu}, M_{\lam\s} \right] &=& 
        -i \left( \eta_{\mu\lam}M_{\nu\s} + \eta_{\nu\s}M_{\mu\lam} 
                  - \eta_{\mu\s}M_{\nu\lam} - \eta_{\nu\lam}M_{\mu\s} \right)
          \nonumber \\
    \left[ D, P_\mu \right] &=& -i P_\mu , \quad \left[ D, M_{\mu\nu} \right] = 0, 
             \quad \left[ D, K_\mu \right] = i K_\mu , 
          \nonumber \\
    \left[ M_{\mu\nu}, K_\lam \right] &=& -i \left( \eta_{\mu\lam} K_\nu - \eta_{\nu\lam} K_\mu \right),
             \quad \left[ K_\mu, K_\nu \right] = 0 ,
          \nonumber \\
    \left[ K_\mu, P_\nu \right] &=& 2i \left( \eta_{\mu\nu}D + M_{\mu\nu} \right) .
\end{eqnarray}

The gauge ghost field $c^\mu$ satisfying conformal Killing equations is expanded by 15 Grassmannian modes as
\begin{eqnarray}
   c^\mu(x) = \rc_-^\mu   + 2 x_\nu \rc^{\nu\mu}  + x^\mu \rc   + x^2 \rc_+^\mu  - 2 x^\mu x_\nu \rc_+^\nu ,
\end{eqnarray}
where $\rc_-^\mu$, $\rc^{\mu\nu}$, $\rc$, $\rc_+^\mu$ are Hermitian operators and $\rc^{\mu\nu}$ is antisymmetric. The $\rc$ and $\rc^{\mu\nu}$ modes are dimensionless, while $\rc_-^\mu$ and $\rc_+^\mu$ have dimensions $-1$ and $1$, respectively.

We also introduce the 15 antighost modes $\rb_-^\mu$, $\rb^{\mu\nu}$, $\rb$ and $\rb_+^\mu$ with the same properties that the gauge ghost modes have. The anticommutation relations between gauge ghosts and antighosts are set as
\begin{eqnarray}
      \left\{ \rc_-^\mu , \rb_+^\nu \right\} &=& \left\{ \rc_+^\mu , \rb_-^\nu \right\} = \eta^{\mu\nu} ,
      \quad \left\{ \rc , \rb \right\} = 1 , 
        \nonumber \\
    \left\{ \rc^{\mu\nu} , \rb^{\lam\s} \right\} &=& \eta^{\mu\lam} \eta^{\nu\s} - \eta^{\mu\s} \eta^{\nu\lam} .
\end{eqnarray}
The generators of conformal algebra in the gauge ghost sector are given by
\begin{eqnarray}
    P_\gh^\mu &=& i \left( - 2 \rb \rc_+^\mu  + \rb_+^\mu \rc   + \rb^\mu_{~\lam} \rc_+^\lam  
                           + 2 \rb_+^\lam \rc^\mu_{~\lam}   \right) ,
                    \nonumber \\
    M_\gh^{\mu\nu} &=& i \left( \rb_+^\mu \rc_-^\nu  - \rb_+^\nu \rc_-^\mu   
                                + \rb_-^\mu \rc_+^\nu  - \rb_-^\nu \rc_+^\mu
                                + \rb^{\mu\lam} \rc^\nu_{~\lam}  - \rb^{\nu\lam} \rc^\mu_{~\lam}  \right) ,
                    \nonumber \\
    D_\gh &=& i \left( \rb_-^\lam \rc_{+ \lam}   - \rb_+^\lam \rc_{- \lam}  \right), 
                    \nonumber \\
    K^\mu_\gh &=& i \left(  2 \rb \rc_-^\mu  - \rb_-^\mu \rc   + \rb^\mu_{~\lam} \rc_-^\lam  
                           + 2 \rb_-^\lam \rc^\mu_{~\lam}   \right) . 
\end{eqnarray}
The nilpotent BRST operator is then defined by 
\begin{eqnarray}
     Q_\BRST &=& \rc_-^\mu \left( P_\mu + \half P^\gh_\mu \right) 
               + \rc^{\mu\nu} \left( M_{\mu\nu} + \half M^\gh_{\mu\nu} \right) 
               + \rc \left( D + \half D^\gh \right) 
                  \nonumber \\
             &&  + \rc_+^\mu \left( K_\mu + \half K^\gh_\mu \right) 
                  \nonumber \\
             &=& \rc \left( D + D^\gh \right) + \rc^{\mu\nu} \left( M_{\mu\nu} + M^\gh_{\mu\nu} \right)
              - \rb N - \rb^{\mu\nu} N_{\mu\nu} + {\hat Q} ,
\end{eqnarray}
where
\begin{eqnarray}
    N &=& 2 i \rc_+^\mu \rc_{- \mu} , 
          \quad
    N^{\mu\nu} = \fr{i}{2} \left( \rc_+^\mu \rc_-^\nu + \rc_-^\mu \rc_+^\nu \right) 
                   + i \rc^{\mu\lam} \rc^\nu_{~\lam} ,
            \nonumber \\
    {\hat Q} &=& \rc_-^\mu P_\mu + \rc_+^\mu K_\mu .
\end{eqnarray}
The generators $P_\mu$, $M_{\mu\nu}$, $D$ and $K_\mu$ are defined in (\ref{Poincare generator}), (\ref{dilatation generator}) and (\ref{special conformal transformation generator}).

The conformal transformations for Riegert and gauge ghost fields are then expressed as $i [ Q_\BRST, \phi ] = \dl_{\rm B} \phi$ and $i \{ Q_\BRST, c^\mu \} = \dl_{\rm B} c^\mu$. In the following, we construct various physical fields that are invariant under the BRST transformation.

\section{Cosmological Constant Term}
\setcounter{equation}{0}
\noindent

First of all, we reconsider the physical operator corresponding to the cosmological constant term $\sq{-g}$. It is represented by the most simple conformal field given by the exponential function of the Riegert field defined as
\begin{equation}
      \calV_\a (x) = \l: e^{\a\phi(x)} \r: = \sum_{n=0}^\infty \fr{\a^n}{n!} \l: \phi^n(x) \r: .
           \label{conformal field V_a}
\end{equation}
The physical field is a conformal field with definite constant $\a$ determined by the BRST invariance condition below. The constant $\a$, called the Riegert charge, includes quantum corrections, which is a real number reflecting that the physical field is a gravitational quantity.

It was shown in \cite{hamada12a} that the conformal field $\calV_\a$ transforms as a scalar under conformal transformations:
\begin{eqnarray}
      i\left[ P_\mu, \calV_\a(x) \right] &=& \pd_\mu \calV_\a(x) ,
             \nonumber \\
   i \left[ M_{\mu\nu}, \calV_\a(x) \right] &=& \left( x_\mu \pd_\nu - x_\nu \pd_\mu \right) \calV_\a (x) ,
             \nonumber \\ 
   i\left[ D, \calV_\a(x) \right] &=& \left( x^\mu \pd_\mu + h_\a \right) \calV_\a(x) ,
             \nonumber \\
   i \left[ K_\mu, \calV_\a (x) \right] 
       &=& \left( x^2 \pd_\mu - 2x_\mu x^\nu \pd_\nu -2 x_\mu h_\a \right) \calV_\a (x) ,
            \label{conformal transformation of V_a}
\end{eqnarray}
where the conformal dimension is given by $h_\a$ calculated to be
\begin{equation}
     h_\a = \a -\fr{\a^2}{4b_1} .
        \label{conformal dimension h_a}
\end{equation}
The second term proportional to $1/b_1$ is the quantum correction.

From these equations, we find that the BRST transformation of the conformal field $\calV_\a$ is given by 
\begin{eqnarray}
    i \left[ Q_\BRST , \calV_\a(x) \right] = c^\mu\pd_\mu \calV_\a (x) + \fr{h_\a}{4} \pd_\mu c^\mu \calV_\a(x) ,
\end{eqnarray}
Thus, the first example of the physical field is given by the conformal field with definite conformal dimension $h_\a =4$ such that
\begin{eqnarray}
    i \left[ Q_\BRST, \int d^4 x \calV_\a(x) \right] = \int d^4 x \pd_\mu \{ c^\mu \calV_\a(x) \} = 0 .
            \label{physical condition I}
\end{eqnarray}

The physical field can be made BRST invariant locally by introducing the function of gauge ghosts defined by 
\begin{equation}
     \om = \fr{1}{4!} \eps_{\mu\nu\lam\s}c^\mu c^\nu c^\lam c^\s .
          \label{omega}
\end{equation}
This operator transforms as 
\begin{equation}
    i \left[ Q_\BRST, \om(x) \right] = c^\mu \pd_\mu \om(x) = -\om \pd_\mu c^\mu(x) ,
\end{equation} 
where $c^\mu \om=0$ is used. Using this commutator, we can show that the operator product $\om \calV_\a$ becomes BRST invariant without the spacetime volume integral as
\begin{equation}
     i \left[ Q_\BRST , \om \calV_\a(x) \right] = \fr{1}{4} \left( h_\a -4 \right) \om \pd_\mu c^\mu \calV_\a(x) = 0 
\end{equation}
for $h_\a =4$.

There are two solutions for the equation $h_\a =4$. We select the Riegert charge 
\begin{eqnarray}
    \a = 2b_1 \left( 1 - \sq{1-\fr{4}{b_1}}  \right)
        \label{Riegert charge a}
\end{eqnarray}  
for the cosmological constant term because it approaches the canonical value $4$ in the classical limit $b_1 \to \infty$ corresponding to the large number limit of matter fields coupled to gravity. In the following, $\a$ is fixed to be the value (\ref{Riegert charge a}) if not specified.

The another solution of $h_\a=4$ is given by $4b_1-\a$ due to the duality relation $h_\a = h_{4b_1-\a}$. The operator $\calV_{4b_1-\a}$ does not reduce to the canonical form of the cosmological constant term at the classical limit, but this operator is regarded as the adjoint of $\calV_\a$ in the presence of the background charge as discussed in Section 6.

The physical condition (\ref{physical condition I}) is equivalent to the condition that the physical field commutes with all generators of conformal symmetry, $P_\mu$, $M_{\mu\nu}$, $D$ and $K_\mu$. In general, physical fields are given by scalar fields with conformal dimension four, while fields with tensor indices do not satisfy the physical condition due to the presence of spin terms in their conformal transformations.

\section{Ricci Scalar Curvature}
\setcounter{equation}{0}
\noindent

Next, we consider the most simple diffeomorphism invariant scalar operator with derivatives, namely the Ricci scalar curvature. It was first constructed on the $R \times S^3$ background in \cite{hamada12b}. In this section, we will construct it on $M^4$. The result is consistent with that of $R \times S^3$.

The quantum Ricci scalar operator will be composed of four second-order operators $\l: \P_\chi \phi^n \r:$, $\l: \lap3 \phi \phi^n \r:$, $\l: \chi^2 \phi^n \r:$ and $\l: \pd_k \phi \pd^k \phi \phi^n \r:$ (which are the special cases of $\Phi_n^{a,b,c,d}$ (\ref{definition of Phi_n}) discussed in the next section). So, we seek a combination of these four operators that has a good behavior under conformal transformations.

From the requirement that the operator transforms as a scalar under Lorentz transformation, we consider the following two combinations:
\begin{eqnarray}
    \calR_n^1 &=&  \l: \pd^2 \phi \phi^n \r: 
               ~=~ \l: \left( \fr{4\pi^2}{b_1} \P_\chi + \lap3 \phi \right) \phi^n \r: ,
                  \nonumber \\  
    \calR_n^2 &=&  \l: \pd_\lam \phi \pd^\lam \phi \phi^n \r: 
               ~=~ \l: \left( -\chi^2 + \pd_k \phi \pd^k \phi \right) \phi^n \r: .
\end{eqnarray}
Under dilatations and special conformal transformations, we find that these combinations transform as
\begin{eqnarray}
   i \left[ D, \calR_n^{1,2}(x) \right] = \left( x^\mu \pd_\mu + 2 \right) \calR_n^{1,2}(x)
             + n \calR_{n-1}^{1,2}(x) - \fr{1}{4b_1} n(n-1) \calR_{n-2}^{1,2}(x) 
         \label{dilataion of R_n^{1,2}}
\end{eqnarray}
and
\begin{eqnarray}
   i \left[ K_\mu, \calR_n^1(x) \right] 
      &=& \left\{  x^2 \pd_\mu - 2 x_\mu x^\lam \pd_\lam - 4 x_\mu \right\} \calR_n^1(x)
             \nonumber \\
       &&  - 2 x_\mu \left( n \calR_{n-1}^1(x) - \fr{1}{4b_1} n(n-1) \calR_{n-2}^1(x) \right)
             \nonumber \\
       &&  + 4 \l: \pd_\mu \phi \phi^n(x) \r: ,
               \nonumber \\
   i \left[ K_\mu, \calR_n^2(x) \right] 
       &=& \left\{  x^2 \pd_\mu - 2 x_\mu x^\lam \pd_\lam - 4 x_\mu \right\} \calR_n^2(x)
             \nonumber \\
       &&  - 2 x_\mu \left( n \calR_{n-1}^2(x) - \fr{1}{4b_1} n(n-1) \calR_{n-2}^2(x) \right)
             \nonumber \\
       &&  - 4 \l: \pd_\mu \phi \phi^n(x) \r: 
           + \fr{1}{b_1} n  \l: \pd_\mu \phi \phi^{n-1}(x) \r: .
\end{eqnarray}
Here, the terms with $1/b_1$ in the $D$ and $K_\mu$ transformations are quantum corrections.

In order to put together quantum correction terms, we consider the following exponentiated operators:
\begin{eqnarray}
     \calR^{1,2}_\b =  \sum_{n=0}^\infty \fr{\b^n}{n!} \calR_n^{1,2} .
\end{eqnarray}
From (\ref{dilataion of R_n^{1,2}}), each operator transforms as a scalar field with conformal dimension $h_\b +2$ under dilatations as follows:
\begin{eqnarray}
   i \left[ D, \calR_\b^{1,2}(x) \right] = \left( x^\mu \pd_\mu + h_\b + 2 \right) \calR_\b^{1,2}(x) ,
\end{eqnarray}
where $h_\b$ is defined by (\ref{conformal dimension h_a}). For special conformal transformations, they transform as
\begin{eqnarray}
   i \left[ K_\mu, \calR^1_\b(x) \right] 
     &=&  \left\{  x^2 \pd_\mu - 2 x_\mu x^\lam \pd_\lam - 2 x_\mu \left( h_\b + 2 \right) \right\} \calR^1_\b(x) 
         + 4 \l: \pd_\mu \phi e^{\b \phi}(x) \r: ,   
             \nonumber \\       
   i \left[ K_\mu, \calR^2_\b(x) \right] 
     &=&  \left\{  x^2 \pd_\mu - 2 x_\mu x^\lam \pd_\lam - 2 x_\mu \left( h_\b + 2 \right) \right\} \calR^2_\b(x) 
             - 4 \fr{h_\b}{\b} \l: \pd_\mu \phi e^{\b \phi}(x) \r: .
              \nonumber \\
      &&   \label{K_mu for R^1 and R^2}
\end{eqnarray}
Here, the last term in each transformation breaks that the field behaves as a conformal field.

Therefore, we consider the combination
\begin{eqnarray}
    \calR_\b = \calR^1_\b + \fr{\b}{h_\b} \calR^2_\b 
      = \l: e^{\b\phi} \left( \pd^2 \phi + \fr{\b}{h_\b} \pd_\lam \phi \pd^\lam \phi \right) \r: .
         \label{quantum Ricci scalar}
\end{eqnarray}
We then find that it transforms under conformal transformations as
\begin{eqnarray}
  i \left[ P_\mu, \calR_\b(x) \right] &=&  \pd_\mu \calR_\b(x) ,
            \nonumber \\
  i \left[ M_{\mu\nu}, \calR_\b(x) \right] &=&  \left( x_\mu \pd_\nu - x_\nu \pd_\mu \right) \calR_\b(x) ,
            \nonumber \\   
  i \left[ D, \calR_\b(x) \right] 
     &=&  \left( x^\lam \pd_\lam + h_\b +2 \right) \calR_\b(x) ,
            \nonumber \\
  i \left[ K_\mu, \calR_\b(x) \right] 
     &=&  \left\{  x^2 \pd_\mu - 2 x_\mu x^\lam \pd_\lam - 2 x_\mu \left(h_\b + 2 \right) \right\} \calR_\b(x) .
\end{eqnarray}
Thus, $\calR_\b$ transforms as a scalar field with conformal dimension $h_\b + 2$ and the BRST transformation is given by
\begin{eqnarray}
    i \left[ Q_\BRST , \calR_\b(x) \right] = c^\mu\pd_\mu \calR_\b (x) + \fr{h_\b +2}{4} \pd_\mu c^\mu \calR_\b(x) .
\end{eqnarray}

Therefore, the spacetime volume integral of $\calR_\b$ with $h_\b=2$ becomes BRST invariant such that 
\begin{eqnarray}
   i \left[ Q_\BRST, \int d^4 x \calR_\b(x) \right]
     = \int d^4 x \pd_\mu \left\{ c^\mu \calR_\b(x) \right\} = 0 .
          \label{BRST condition for Ricci scalar 1}
\end{eqnarray}
The quantum Ricci scalar curvature is now identified to be $\calR_\b$ with the Riegert charge
\begin{eqnarray}
   \b = 2b_1 \left( 1- \sq{1-\fr{2}{b_1}} \right) ,
\end{eqnarray}
which is one of the solution of $h_\b =2$. The field $\calR_\b$ then reduces to the canonical form of Ricci scalar curvature $\sq{-g}R$ divided by $-6$ when we take the classical limit $b_1 \to \infty$ such that $\b \to 2$ and $\b/h_\b \to 1$. In the following, $\b$ is fixed to be this value.

As discussed in Section 3, using the operator of gauge ghosts $\om$ (\ref{omega}), we can make the physical field locally BRST invariant as 
\begin{eqnarray}
   \left[ Q_\BRST, \om \calR_\b(x) \right] = 0 . 
        \label{BRST condition for Ricci scalar 2}
\end{eqnarray}
Here, note that this BRST invariance condition is stronger than the condition (\ref{BRST condition for Ricci scalar 1}) because the condition (\ref{BRST condition for Ricci scalar 1}) holds up to total divergences. Indeed, both of the fields $\calR^1_\b$ and $\calR^2_\b$ satisfy the BRST invariance condition (\ref{BRST condition for Ricci scalar 1}) because the last terms in (\ref{K_mu for R^1 and R^2}) that break conformally covariant behaviors are totally divergent, but they do not satisfy the condition (\ref{BRST condition for Ricci scalar 2}).

\section{Integer Power of Ricci Scalar Curvature}
\setcounter{equation}{0}
\noindent

Let us generalize the argument to higher derivative fields. From the expression of quantum Ricci scalar curvature (\ref{quantum Ricci scalar}), we guess that the physical field corresponding to the $m$-th power of Ricci scalar will be given by the following composite operator:
\begin{eqnarray}
    \calP_\gm^{[m]} = \sum_{n=0}^\infty \fr{\gm^n}{n!} \calP_n^{[m]} 
      = \l: e^{\gm\phi} \left( \pd^2 \phi + \fr{\gm}{h_\gm} \pd_\lam \phi \pd^\lam \phi \right)^m \r: ,
           \label{integer power of Ricci scalar}
\end{eqnarray}
where 
\begin{eqnarray}
   \calP_n^{[m]} &=& \l: \left( \pd^2 \phi + \fr{\gm}{h_\gm} \pd_\lam \phi \pd^\lam \phi \right)^m \phi^n \r:
            \nonumber \\
       &=& \sum_{a=0}^m  \Bigl(\!\!\begin{array}{c} m \\ a \end{array}\!\!\Bigr)
                        \left( \fr{\gm}{h_\gm} \right)^{m-a} 
                        \l: \left( \pd^2 \phi \right)^a \left( \pd_\lam \phi \pd^\lam \phi \right)^{m-a} \phi^n \r:
            \nonumber \\
       &=& \sum_{a=0}^m \sum_{b=0}^a \sum_{c=0}^{m-a} 
            \Bigl(\!\!\begin{array}{c} m \\ a \end{array}\!\!\Bigr)
            \Bigl(\!\!\begin{array}{c} a \\ b \end{array}\!\!\Bigr)
            \Bigl(\!\!\begin{array}{c} m-a \\ c \end{array}\!\!\Bigr)
            \left( \fr{\gm}{h_\gm} \right)^{m-a} \left( \fr{4\pi^2}{b_1} \right)^b (-1)^c
            \Phi_n^{b,a-b,c,m-a-c} .
              \nonumber \\
       &&  \label{definition of P_n^[m]}
\end{eqnarray}
The last operator is defined by 
\begin{eqnarray}
    \Phi_n^{a,b,c,d} 
    = \l: \P_\chi^a \left( \lap3 \phi \right)^b \chi^{2c} \left( \pd_k \phi \pd^k \phi \right)^d \phi^n \r: ,
           \label{definition of Phi_n}
\end{eqnarray}
where $a$, $b$, $c$, and $d$ are non-negative integers. In the following, we will show that the field $\calP_\gm^{[m]}$ with appropriate Riegert charge $\gm$ becomes BRST invariant.

We first present the transformation law of $\Phi_n^{a,b,c,d}$ calculated using the results given in Appendix B. For translations and spatial rotations, we obtain
\begin{eqnarray}
   i \left[ P_\mu,  \Phi_n^{a,b,c,d}(x) \right] 
       &=& \pd_\mu \Phi_n^{a,b,c,d}(x) ,
               \nonumber \\
   i \left[ M_{ij},  \Phi_n^{a,b,c,d}(x) \right] 
       &=& \left( x_i \pd_j - x_j \pd_i \right) \Phi_n^{a,b,c,d}(x) ,
\end{eqnarray}
while the Lorentz boosts have extra terms as
\begin{eqnarray}
   i \left[ M_{0j},  \Phi_n^{a,b,c,d}(x) \right] 
       &=& -\left( \eta \pd_j + x_j \pd_\eta \right) \Phi_n^{a,b,c,d}(x) 
           +2a \fr{b_1}{4\pi^2} \l: \pd_j \chi \Phi_n^{a-1,b,c,d}(x) \r: 
                  \nonumber \\
       &&  -2b \l: \pd_j \chi \Phi_n^{a,b-1,c,d}(x) \r:
           -2c \l: \chi \pd_j \phi \Phi_n^{a,b,c-1,d}(x) \r: 
                  \nonumber \\   
       &&  -2d \l: \chi \pd_j \phi \Phi_n^{a,b,c,d-1}(x) \r: .
\end{eqnarray}
The last four terms in the right-hand side come from the fact that $\Phi_n^{a,b,c,d}$ is not a Lorentz scalar. Since these terms are not quantum corrections, they cancel out when we consider combinations of Lorentz scalar at the classical level. Thus, we show that the operator $\calP^{[m]}_\gm$ transforms as a scalar field under Lorentz transformations.

Dilatations are given by
\begin{eqnarray}
    i \left[ D, \Phi_n^{a,b,c,d}(x) \right] 
      &=& \left\{ x^\mu \pd_\mu + 2(a+b+c+d) \right\} \Phi_n^{a,b,c,d}(x)
              \nonumber \\
       && + n \Phi_{n-1}^{a,b,c,d}(x) - \fr{1}{4b_1} n(n-1) \Phi_{n-2}^{a,b,c,d}(x) ,
          \label{commutator of D and Phi_n}
\end{eqnarray}
where the last term with $1/b_1$ is a quantum correction. For special conformal transformations, we find
\begin{eqnarray}
    i \left[ K_0, \Phi_n^{a,b,c,d}(x) \right] 
      &=& \left\{ (\eta^2 + \bx^2) \pd_\eta + 2\eta x^k \pd_k +4(a+b+c+d)\eta \right\} \Phi_n^{a,b,c,d}(x)
              \nonumber \\
       && + 2\eta \left( n \Phi_{n-1}^{a,b,c,d}(x) - \fr{1}{4b_1} n(n-1) \Phi_{n-2}^{a,b,c,d}(x) \right)
              \nonumber \\
       && + 4 x^k \biggl( - a \fr{b_1}{4\pi^2}\l: \pd_k \chi \Phi_n^{a-1,b,c,d}(x) \r:
                         + b \l: \pd_k \chi \Phi_n^{a,b-1,c,d}(x) \r:
              \nonumber \\
      &&  \qquad\quad
                         + c \l: \chi \pd_k \phi \Phi_n^{a,b,c-1,d}(x) \r:
                         + d \l: \chi \pd_k \phi \Phi_n^{a,b,c,d-1}(x) \r: \biggr)
              \nonumber \\
      && - 2a \fr{b_1}{4\pi^2} \l: \chi \Phi_n^{a-1,b,c,d}(x) \r:
         + 6b \l: \chi \Phi_n^{a,b-1,c,d}(x) \r:
              \nonumber \\
      && + 4c  \l: \chi \Phi_n^{a,b,c-1,d}(x) \r:
         - c n \fr{1}{b_1} \l: \chi \Phi_{n-1}^{a,b,c-1,d}(x) \r: 
              \label{commutator of K_0 and Phi_n 1}
\end{eqnarray}
and
\begin{eqnarray}
    i \left[ K_j, \Phi_n^{a,b,c,d}(x) \right] 
      &=& \Bigl\{ (-\eta^2 + \bx^2) \pd_j - 2x_j x^k \pd_k - 2\eta x_j \pd_\eta 
              \nonumber \\
       && \qquad\qquad\qquad\quad
                 - 4(a+b+c+d)x_j  \Bigr\} \Phi_n^{a,b,c,d}(x)
              \nonumber \\
       && - 2x_j \left( n \Phi_{n-1}^{a,b,c,d}(x) - \fr{1}{4b_1} n(n-1) \Phi_{n-2}^{a,b,c,d}(x) \right)
              \nonumber \\
       && - 4 \eta \biggl( - a \fr{b_1}{4\pi^2}\l: \pd_j \chi \Phi_n^{a-1,b,c,d}(x) \r:
                         + b \l: \pd_j \chi \Phi_n^{a,b-1,c,d}(x) \r:
              \nonumber \\
      &&  \qquad\quad
                         + c \l: \chi \pd_j \phi \Phi_n^{a,b,c-1,d}(x) \r:
                         + d \l: \chi \pd_j \phi \Phi_n^{a,b,c,d-1}(x) \r: \biggr)
              \nonumber \\
      && + 2a \fr{b_1}{4\pi^2} \l: \pd_j \phi \Phi_n^{a-1,b,c,d}(x) \r:
         + 2b \l: \pd_j \phi \Phi_n^{a,b-1,c,d}(x) \r:
              \nonumber \\
      && - 4d  \l: \pd_j \phi \Phi_n^{a,b,c,d-1}(x) \r:
         + d n \fr{1}{b_1} \l: \pd_j \phi \Phi_{n-1}^{a,b,c,d-1}(x) \r: .
              \label{commutator of K_j and Phi_n}
\end{eqnarray}
Here, the four terms with $1/b_1$ in the $K_0$ and $K_j$ transformations are quantum corrections.\footnote{ 
Note that the terms with $b_1/4\pi^2$ are not quantum corrections.
}  

From (\ref{commutator of D and Phi_n}), we find that the exponentiated operator
\begin{eqnarray}
   \Phi_\gm^{a,b,c,d}= \sum_{n=0}^\infty \fr{\gm^n}{n!} \Phi_n^{a,b,c,d} 
         \label{definition of Phi_gm}
\end{eqnarray}
transforms under dilatations as a field with conformal dimension $h_\gm+2(a+b+c+d)$ such that
\begin{eqnarray}
     i \left[ D, \Phi_\gm^{a,b,c,d}(x) \right] 
      = \left\{ x^\mu \pd_\mu + h_\gm + 2(a+b+c+d) \right\} \Phi_\gm^{a,b,c,d}(x) ,
            \label{dilatation of Phi_gm}
\end{eqnarray}
where $h_\gm$ is defined by (\ref{conformal dimension h_a}). From this, we find that the operator $\calP^{[m]}_\gm$ transforms as a scalar field with conformal dimension $h_\gm +2m$ under dilatations.

As for special conformal transformations, the situation becomes more complicated. In the following, we further study the transformation laws for the combination $\calP^{[m]}_n$. From (\ref{commutator of K_0 and Phi_n 1}), we find that $\Phi_n^{b,a-b,c,m-a-c}$ in (\ref{definition of P_n^[m]}) transforms under $K_0$ as
\begin{eqnarray}
   && i \left[ K_0 , \Phi_n^{b,a-b,c,m-a-c} \right]
            \nonumber \\
   && = \left\{ (\eta^2 + \bx^2) \pd_\eta + 2\eta x^k \pd_k + 4m\eta \right\} \Phi_n^{b,a-b,c,m-a-c}
            \nonumber \\
   && \quad
       + 2\eta \left\{ n \Phi_{n-1}^{b,a-b,c,m-a-c} 
                      -\fr{1}{4b_1} n(n-1) \Phi_{n-2}^{b,a-b,c,m-a-c} \right\}
              \nonumber \\
   && \quad
       + 4 x^k \biggl\{ -b \fr{b_1}{4\pi^2} \l: \pd_k \chi \Phi_n^{b-1,a-b,c,m-a-c} \r:
                       + (a-b) \l: \pd_k \chi \Phi_n^{b,a-b-1,c,m-a-c} \r:
              \nonumber \\
   && \qquad
                 + c \l: \chi \pd_k \phi \Phi_n^{b,a-b,c-1,m-a-c} \r:
                 + (m-a-c) \l: \chi \pd_k \phi \Phi_n^{b,a-b,c,m-a-c-1} \r:  \biggr\}
             \nonumber \\
   && \quad
        -2b \fr{b_1}{4\pi^2} \l: \chi \Phi_n^{b-1,a-b,c,m-a-c} \r:
        +6(a-b) \l: \chi \Phi_n^{b,a-b-1,c,m-a-c} \r:
              \nonumber \\
    && \quad
         +4c \l: \chi \Phi_n^{b,a-b,c-1,m-a-c} \r: 
         - \fr{1}{b_1} n c \l: \chi \Phi_{n-1}^{b,a-b,c-1,m-a-c} \r: .
              \label{commutator of K_0 and Phi_n 2}
\end{eqnarray}
Here, taking the sum over $b$ in (\ref{definition of P_n^[m]}) and taking into account the relations
\begin{eqnarray}
    b \Bigl(\!\!\begin{array}{c} a \\ b \end{array}\!\!\Bigr) 
    = a \Bigl(\!\!\begin{array}{c} a-1 \\ b-1 \end{array}\!\!\Bigr), 
        \qquad
    (a-b) \Bigl(\!\!\begin{array}{c} a \\ b \end{array}\!\!\Bigr)
    = a \Bigl(\!\!\begin{array}{c} a-1 \\ b \end{array}\!\!\Bigr), 
\end{eqnarray}
we find that the first two terms in the braces multiplied by $4x^k$ cancel out. The third and fourth terms in this braces also cancel out in the same way when we take the sum over $c$. Thus, the terms multiplied by $4x^k$ give no contributions.

The last four terms in (\ref{commutator of K_0 and Phi_n 2}) give finite contributions. We here evaluate the one of them as follows:
\begin{eqnarray}
    &&  \sum_{a=0}^m \sum_{b=0}^a \sum_{c=0}^{m-a} 
            \Bigl(\!\!\begin{array}{c} m \\ a \end{array}\!\!\Bigr)
            \Bigl(\!\!\begin{array}{c} a \\ b \end{array}\!\!\Bigr)
            \Bigl(\!\!\begin{array}{c} m-a \\ c \end{array}\!\!\Bigr)
            \left( \fr{\gm}{h_\gm} \right)^{m-a} \left( \fr{4\pi^2}{b_1} \right)^b (-1)^c
               \nonumber \\
    && \qquad \times
          (-2b) \fr{b_1}{4\pi^2} \l: \chi \Phi_n^{b-1,a-b,c,m-a-c} \r:
              \nonumber \\
    && = -2m \sum_{a=1}^{m} \sum_{b=1}^a \sum_{c=0}^{m-a} 
             \Bigl(\!\!\begin{array}{c} m-1 \\ a-1 \end{array}\!\!\Bigr)
             \Bigl(\!\!\begin{array}{c} a-1 \\ b-1 \end{array}\!\!\Bigr)
             \Bigl(\!\!\begin{array}{c} m-a \\ c \end{array}\!\!\Bigr)
             \left( \fr{\gm}{h_\gm} \right)^{m-a}  
               \nonumber \\
     && \qquad\qquad \times 
            \left( \fr{4\pi^2}{b_1} \right)^{b-1} (-1)^c \l: \chi \Phi_n^{b-1,a-b,c,m-a-c} \r:
               \nonumber \\
    && = -2m \sum_{a=0}^{m-1} \sum_{b=0}^a \sum_{c=0}^{m-1-a} 
             \Bigl(\!\!\begin{array}{c} m-1 \\ a \end{array}\!\!\Bigr)
             \Bigl(\!\!\begin{array}{c} a \\ b \end{array}\!\!\Bigr)
             \Bigl(\!\!\begin{array}{c} m-1-a \\ c \end{array}\!\!\Bigr)
             \left( \fr{\gm}{h_\gm} \right)^{m-1-a}  
               \nonumber \\
     && \qquad\qquad \times 
            \left( \fr{4\pi^2}{b_1} \right)^b (-1)^c \l: \chi \Phi_n^{b,a-b,c,m-1-a-c} \r:
               \nonumber \\
     && = -2m \l: \chi \calP_n^{[m-1]} \r: .
\end{eqnarray}
The other three terms are also evaluated in the same way.

Thus, we find that $\calP^{[m]}_n$ has the following transformation law:
\begin{eqnarray}
    i \left[ K_0 , \calP_n^{[m]} \right]
    &=& \left\{ (\eta^2 + \bx^2) \pd_\eta + 2\eta x^k \pd_k + 4m\eta \right\} \calP_n^{[m]}
              \nonumber \\
     && + 2\eta \left( n \calP_{n-1}^{[m]} -\fr{1}{4b_1} n(n-1) \calP_{n-2}^{[m]} \right)
              \nonumber \\
     && + 4m \left\{ \left( 1 - \fr{\gm}{h_\gm} \right) \l: \chi \calP_n^{[m-1]} \r: 
        + \fr{1}{4b_1} \fr{\gm}{h_\gm} n \l: \chi \calP_{n-1}^{[m-1]} \r: \right\} .
              \nonumber \\
     &&   \label{commutator of K_0 and P_gm^[m]}
\end{eqnarray}
Similarly, the transformation law under $K_j$ is calculated as
\begin{eqnarray}
   i \left[ K_j , \calP_n^{[m]} \right]
    &=& \left\{ (-\eta^2 + \bx^2) \pd_j - 2x_j x^k \pd_k -2\eta x_j \pd_\eta - 4m x_j \right\} \calP_n^{[m]}
              \nonumber \\
    &&  - 2x_j \left( n \calP_{n-1}^{[m]} -\fr{1}{4b_1} n(n-1) \calP_{n-2}^{[m]} \right)
              \nonumber \\
    &&  + 4m \left\{ \left( 1 - \fr{\gm}{h_\gm} \right) \l: \pd_j \phi \calP_n^{[m-1]} \r: 
        + \fr{1}{4b_1} \fr{\gm}{h_\gm} n \l: \pd_j \phi \calP_{n-1}^{[m-1]} \r: \right\} .
              \nonumber \\
    &&     \label{commutator of K_j and P_gm^[m]}
\end{eqnarray}

Putting together quantum correction terms in the exponentiated form (\ref{integer power of Ricci scalar}), we find that both of the last braces multiplied by $4m$ in (\ref{commutator of K_0 and P_gm^[m]}) and (\ref{commutator of K_j and P_gm^[m]}) give vanishing contributions due to $1-\gm/h_\gm + \gm^2/4b_1 h_\gm = 0$. Thus, we obtain the desired transformation law 
\begin{eqnarray}
    i \left[ K_\mu , \calP_\gm^{[m]}(x) \right]
     &=& \left\{ x^2 \pd_\mu - 2 x_\mu x^\nu \pd_\nu - 2x_\mu ( h_\gm + 2m) \right\} \calP_\gm^{[m]}(x) .
\end{eqnarray}
In this way, we find that $\calP_\gm^{[m]}$ transforms as a scalar field with conformal dimension $h_\gm + 2m$ and the BRST transformation law is given by
\begin{eqnarray}
    i \left[ Q_\BRST , \calP_\gm^{[m]}(x) \right] 
    = c^\mu\pd_\mu \calP_\gm^{[m]} (x) + \fr{h_\gm +2m}{4} \pd_\mu c^\mu \calP_\gm^{[m]}(x) .
\end{eqnarray}

Therefore, the BRST invariant physical field is given by $\calP_\gm^{[m]}$ with $h_\gm = 4 -2m$ such that $[ Q_\BRST, \om \calP_\gm^{[m]}]=0$. The Riegert charge $\gm$ is then determined to
\begin{eqnarray}
   \gm = 2b_1 \left( 1- \sq{1-\fr{4-2m}{b_1}} \right) ,
\end{eqnarray}
which is the solution that the physical field $\calP_\gm^{[m]}$ reduces to the canonical form of $\sq{-g}R^m$ when we take the classical limit $b_1 \to \infty$ such that $\gm \to 4-2m$ and $\gm/h_\gm \to 1$.

\section{Remarks on Correlation Functions}
\setcounter{equation}{0}
\noindent

Let us consider the physical field $\calO_\gm$ with the Riegert charge $\gm$ and the adjoint of it denoted by $\tilde{\calO_\gm}$ satisfying the correlation relation
\begin{eqnarray}
     \langle \tilde{\calO}_\gm(x) \calO_\gm(y) \rangle = \fr{1}{(x-y)^8} . 
\end{eqnarray}
In general, $\tilde{\calO_\gm}$ is not the Hermitian conjugate of $\calO_\gm$, which is now $\calO_\gm$ itself, because the operator product does not produce the identity operator in this case.\footnote{ 
The situation is the same as in the case of the Liouville gravity \cite{seiberg}. If the Riegert charge were purely imaginary such as $\gm=ip$ and there were no background charges, namely, no liner terms of $\phi$ in the action (\ref{Riegert action}), the correlation function between $\calO_{ip}$ and its Hermitian conjugate $\calO_{-ip}$ could exist as in the case of string theory \cite{kato, fms}. 
} 
Here, we discuss this matter with some examples.

The correlation function of the field $\calV_\a$ is evaluated by using the correlator $\langle \phi(x) \phi(y) \rangle = -(1/4b_1) \times \log (x-y)^2$ as
\begin{eqnarray}
    \calV_{\a^\pp}(x) \calV_\a(y) &=&  \exp \left\{ \a^\pp \a \langle \phi(x) \phi(y) \rangle \right\} 
                                \l: \calV_{\a^\pp}(x)\calV_\a(y) \r:
             \nonumber \\
         &=& \left( \fr{1}{(x-y)^2} \right)^{\fr{\a^\pp \a}{4b_1}} \l: \calV_{\a^\pp}(x) \calV_\a(y) \r: ,
\end{eqnarray}
where $\a$ and $\a^\pp$ are taken to be arbitrary. If $\a^\pp$ is taken to be $4b_1-\a$ and $\a$ is fixed to be ({\ref{Riegert charge a}), we obtain  
\begin{eqnarray}
    \calV_{4b_1-\a}(x) \calV_\a(y) = \fr{1}{(x-y)^{8}} \l: \calV_{4b_1-\a}(x) \calV_\a(y) \r: .
\end{eqnarray}
The leading singular term of the right-hand side is given by $\calV_{4b_1}$ which is the identity operator with vanishing conformal weight and its vacuum expectation value is unity. Thus, we obtain
\begin{eqnarray}
    \langle \calV_{4b_1-\a}(x) \calV_\a(y) \rangle = \fr{1}{(x-y)^8} .      
\end{eqnarray}
This implies that $\tilde{\calV}_\a$ is given by $\calV_{4b_1-\a}$, which is another one of the BRST invariant pair derived from the duality relation $h_\a=h_{4b_1-\a}$ as mentioned in Section 3.

In the same way, we obtain the adjoint of $\calR_\b$, which is given by 
\begin{eqnarray}
   \tilde{\calR}_\b = -\fr{b_1}{4} \calR_{4b_1-\b} 
          = - \fr{b_1}{4} \left( \calR^1_{4b_1-\b} + \fr{4b_1-\b}{h_\b} \calR^2_{4b_1-\b} \right)
\end{eqnarray}
such that
\begin{eqnarray}
     \langle \tilde{\calR}_{\b}(x) \calR_\b(y) \rangle = \fr{1}{(x-y)^8} . 
\end{eqnarray}

The field $\tilde{\calR}_\b$ as well as $\tilde{\calV}_\a$ does not reduce to an existing gravitational quantity in the classical limit of $b_1 \to \infty$. In general, $\tilde{\calO}_\gm$ is given by the appropriately normalized field proportional to $\calO_{4b_1-\gm}$, but it has no classical counter quantity. Therefore, it is a purely quantum object that will appear only in the intermediate state of physical correlation functions discussed below.

Now, we are interested in correlation functions among physical fields with correct Riegert charge only. To define them, we should consider the model perturbed by the potential term as in the case of 2D quantum gravity \cite{polyakov, kpz, dk, seiberg, gl}. We here consider the perturbed theory: $S_{\rm RWZ}+ \mu V_\a$ in Wick-rotated Euclidean background, where $\mu$ is the cosmological constant and $V_\a = \int d^4 x \calV_\a$, generally $O_\gm=\int d^4 x \calO_\gm$. At that time, we have to consider the constant mode of the Riegert field $\s$, introduced by the substitution $\phi \to \phi + \s$, and carry out the path integral over the constant mode $A=e^{\a\s}$ first. Noting that $S_{\rm RWZ} \to S_{\rm RWZ} + 4b_1 \s$ and $O_\gm \to A^{\gm/\a} O_\gm$ by the substitution, where the shift term comes from the linear term in the action (\ref{Riegert action}) because the Euler number is two: $(1/32\pi^2 )\int d^4 x \sq{\hg}\hat{G}_4 = 2$. 
We then obtain the correlation function in the perturbed theory as follows:     
\begin{eqnarray}
   \langle\langle O_{\gm_1} \cdots O_{\gm_n} \rangle\rangle 
   &=& \fr{1}{\a} \int^\infty_0 \fr{dA}{A} A^{-s} 
        \langle O_{\gm_1} \cdots O_{\gm_n} e^{-\mu A V_\a} \rangle
         \nonumber \\
   &=& \mu^s \fr{\Gamma(-s)}{\a}  
          \langle O_{\gm_1} \cdots O_{\gm_n} \left( V_\a \right)^s  \rangle ,
\end{eqnarray}
where $s= (4b_1 - \sum_{i=1}^n \gm_i)/\a$. Here, $\langle \cdots \rangle$ represents the correlation function in the unperturbed theory.\footnote{ 
The number $s$ is not integer, but fractional number. Therefore, by regarding $s$ as an integer the correlation function may be evaluated and then $s$ may be analytically continued to the fractional number \cite{gl}. 
} 
This correlation function will exist because the operator product inside the unperturbed correlation function can produce the identity operator.

\section{Conclusion}
\setcounter{equation}{0}
\noindent

We have studied higher derivative physical fields for the background free quantum gravity formulated in terms of a certain CFT on the Minkowski background $M^4$. Since conformal invariance arises as diffeomorphism invariance, the invariance is imposed on the field as well as the vacuum, unlike usual CFT in which the field transforms covariantly, in the context of BRST formalism.

Physical fields are given by scalar fields with conformal dimension four, while tensor fields are excluded. After the quantum cosmological constant term was reinvestigated, we constructed the quantum Ricci scalar field. The results are consistent with those computed on the $R \times S^3$ background as expected from background free nature \cite{hamada12b}. The computation has been generalized to higher derivative fields. In this paper, we have constructed the novel physical field corresponding to the integer power of Ricci scalar curvature.

The BRST invariant fields always appear in pairs due to the presence of the duality in Riegert charges. The physical field was identified with the one that reduces to the existing gravitational quantity in the large $b_1$ limit corresponding to the large number limit of matter fields coupled to gravity.

The Hermitian conjugate of physical field is given by itself reflecting gravitational quantities are real. It is, however, not the adjoint of physical field because the operator product between physical field and itself does not produce the identity operator. The adjoint of physical field is given by another one of the BRST invariant pair, which does not have a classical counterpart so that it is regarded to be a purely quantum object. From this situation, in order to define the correlation function among physical fields with correct Riegert charge, we should consider the model with potential terms such as the cosmological constant term and the Ricci scalar curvature. The correlation function obtained in this way has a power-law behavior for the mass scale appearing in the potential term, as required from cosmological observations \cite{hy, hhy06, hhy10}.

Since conformal symmetry survives as gauge symmetry at the UV limit in this background free model unlike other higher derivative models, the negative-metric mode known as the cause of the unitarity problem can never appear alone because it is not gauge invariant, which is merely one of the elements to make up the physical field. In general, the positivity for the two-point function between a real field and itself seems to be trivial if the path integral is well-defined, namely the action is bounded from below, and our model is this case.\footnote{ 
Of course, this statement is based on the fact that physical fields are written in terms of the Riegert field and thus the negative-metric mode never appears obviously. If negative-metric modes were gauge invariant as in usual perturbative approaches, one could not apply this statement to their two-point functions because in this case only the negative-metric part in the action contributes to the path integral so that the positivity of the overall sign of the action becomes meaningless. 
} 
In the future, we will further study physical properties of correlation functions to give more conclusive statement.

\appendix

\section{Quantum Properties of Riegert Fields}
\setcounter{equation}{0}
\noindent

The Riegert field satisfying the equation of motion $\pd^4 \phi=0$ is expanded by $e^{ik_\mu x^\mu}$ and $\eta e^{ik_\mu x^\mu}$, where $k_\mu x^\mu = - |\bk| \eta + \bk \cdot \bx$. The field is decomposed into the annihilation and creation parts as $\phi = \phi_< + \phi_>$, where $\phi_> = \phi_<^\dag$ and
\begin{equation}
    \phi_<(x) = \fr{\pi}{\sq{b_1}} \int \fr{d^3 \bk}{(2\pi)^{3/2}} \fr{1}{|\bk|^{3/2}}  
                    \left\{ a(\bk) + i|\bk| \eta b(\bk) \right\} e^{ik_\mu x^\mu} .
\end{equation}
The commutation relations for mode operators are then given by the form $[ a(\bk), a^\dag(\bk^\pp) ] = [ a(\bk), b^\dag(\bk^\pp) ] = [ b(\bk), a^\dag(\bk^\pp) ] = \dl_3 (\bk -\bk^\pp)$ and $[ b(\bk), b^\dag(\bk^\pp) ] = 0$.

The operator product of two canonical field variables $A$ and $B$, which are Hermitian operators, is defined by $A(x)B(y) = \lang A(x)B(y) \rang + \l: A(x)B(y) \r:$. The singular part, or the two-point correlation function, is given by $\lang A(x)B(y) \rang = \left[ A_<(x), B_>(y) \right] $, where $A_<$ is the annihilation part of $A$ and $B_>$ is the creation part of $B$. The commutator of $A$ and $B$ can be expressed as $\left[ A(x), B(y) \right] = \lang A(x)B(y) \rang - \lang A(x)B(y) \rang^\dag$.

The two-point correlation function for Riegert field is computed as
\begin{eqnarray}
    \lang \phi(x) \phi(x^\pp) \rang 
          &=& \fr{\pi^2}{b_1} \int_{|\bk| > z} \fr{d^3 \bk}{(2\pi)^3} \fr{1}{|\bk|^3} 
                \left\{ 1 + i|\bk| \left(\eta-\eta^\pp \right) \right\} 
                  e^{ -i|\bk|(\eta -\eta^\pp -i\eps)+ i \bk \cdot (\bx -\bx^\pp) } 
            \nonumber \\  
        &=& -\fr{1}{4b_1} \log \left\{ \left[ -(\eta-\eta^\pp-i\eps)^2 + (\bx -\bx^\pp)^2 \right]  
                                               z^2 e^{2\gm-2} \right\} 
              \nonumber \\
        && -\fr{1}{4b_1}\fr{i\eps}{|\bx-\bx^\pp|}
            \log \fr{\eta-\eta^\pp-i\eps-|\bx-\bx^\pp|}{\eta-\eta^\pp-i\eps+|\bx-\bx^\pp|} .
               \label{phi-phi correlation function}
\end{eqnarray}
Here, $\eps$ is the cutoff parameter to regularize UV divergences and $z$ is an fictitious small mass scale to handle IR divergences.\footnote{ 
In our model, usual mass terms are not gauge invariant and thus the $z$ dependence should cancel out when we consider physical quantities.
} 
At $\eps \to 0$, the correlation function reduces to the well-known form: $\langle \phi(x) \phi(y) \rangle = -(1/4b_1) \times \log (x-y)^2$. The singular parts for other field variables are also obtained by differentiating the correlation function (\ref{phi-phi correlation function}) according to the definitions of field variables such as (\ref{conjugate momentum}). The canonical commutation relation is then expressed, for instance, as $[ \phi(\eta, \bx), \P_\phi(\eta, \bx^\pp) ] = i \eps/\pi^2[(\bx -\bx^\pp)^2 +\eps^2]^2=i\dl_3(\bx-\bx^\pp)$, where the $\dl$ function is regularized as
\begin{eqnarray}
   \dl_3(\bx) = \int \fr{d^3{\bf k}}{(2\pi)^3} e^{i\bk \cdot \bx-\eps \om} = \fr{1}{\pi^2} \fr{\eps}{(\bx^2 +\eps^2)^2} .
         \label{regularized delta function}
\end{eqnarray}

\section{Equal-Time Commutation Relations Among Local Operators}
\setcounter{equation}{0}
\noindent

Here, we summarize the commutation relations between the composite operator $\Phi_n^{a,b,c,d}$ (\ref{definition of Phi_n}) and various local operators that appear in the generators of conformal symmetry. The commutator is evaluated by using the formula
\begin{eqnarray}
  && \left[ \l: AB(x) \r: , \l: \prod_k C_k(y) \r: \right]
          \nonumber \\
   &&= \sum_i \left[ A(x) , C_i(y) \right] \l: B(x)\prod_{k (\neq i)} C_k(y) \r:
       + \sum_i \left[ B(x) , C_i(y) \right] \l: A(x)\prod_{k (\neq i)} C_k(y) \r:
          \nonumber \\
   && \quad
       + \sum_{i,j (i \neq j)}\left\{ \langle A(x)C_i(y) \rangle \langle B(x)C_j(y) \rangle - {\rm H.c.} \right\}
         \l: \prod_{k (\neq i,j)} C_k(y) \r: ,
\end{eqnarray}
where ${\rm H.c.}$ denotes the Hermitian conjugate of the first term in the braces. The last term gives a quantum correction. In the following, we consider the equal-time commutation relations and then the time-coordinate dependence in field variables is disregarded for simplicity.

The equal-time commutation relations with local operators $\calA$ and $\calB_j$ are given by
\begin{eqnarray}
     && i \left[ \calA(\bx), \Phi_n^{a,b,c,d}(\by) \right] 
                 \nonumber \\
     && = \dl_3(\bx-\by) \biggl\{ 
             - a \l: \left(\P_\phi + \fr{b_1}{4\pi^2} \lap3 \chi \right) \Phi_n^{a-1,b,c,d}(\by) \r:
             - 2c \fr{4\pi^2}{b_1} \l: \P_\chi \chi \Phi_n^{a, b, c-1,d}(\by) \r:
                 \nonumber \\
     && \qquad\qquad\qquad
              + n  \l: \chi \Phi_{n-1}^{a,b,c,d}(\by) \r:  \biggr\}
                 \nonumber \\
     && \quad
          + \lap3_x \dl_3 (\bx-\by) \left\{
               - a \fr{b_1}{4\pi^2} \l: \chi(\bx) \Phi_n^{a-1,b,c,d}(\by) \r:
               + b  \l: \chi(\bx) \Phi_n^{a,b-1,c,d}(\by) \r:  \right\}
                 \nonumber \\
     && \quad
           - 2d \pd^k_x \dl_3 (\bx-\by) \l: \chi(\bx) \pd_k \phi \Phi_n^{a,b,c,d-1}(\by) \r: 
                 \nonumber \\
     && \quad
         - a c \fr{20}{\pi^2} f(\bx-\by) \l: \chi \Phi_n^{a-1,b,c-1,d}(\by) \r:
         + b c \fr{80}{b_1} f(\bx-\by) \l: \chi \Phi_n^{a,b-1,c-1,d}(\by) \r:
                 \nonumber \\
     && \quad - c d \fr{4}{b_1} g_k(\bx-\by) \l: \chi \pd^k \phi \Phi_n^{a,b,c-1,d-1}(\by) \r:
\end{eqnarray}
and
\begin{eqnarray}
     && i \left[ \calB_j(\bx), \Phi_n^{a,b,c,d}(\by) \right] 
                 \nonumber \\
     &&= \dl_3(\bx-\by) \left\{ 
              2c \l: \chi \pd_j \chi \Phi_n^{a, b, c-1,d}(\by) \r:
              +n  \l: \pd_j \phi \Phi_{n-1}^{a,b,c,d}(\by) \r:   \right\}
                 \nonumber \\
     && \quad
          + b \lap3_x \dl_3 (\bx-\by) \l: \pd_j \phi(\bx) \Phi_n^{a,b-1,c,d}(\by) \r:
          - a \pd_j^x \dl_3 (\bx-\by) \l: \P_\chi(\bx) \Phi_n^{a-1,b,c,d}(\by) \r: 
                 \nonumber \\ 
     && \quad 
          - 2d \pd^k_x \dl_3 (\bx-\by) \l: \pd_j \phi(\bx) \pd_k \phi \Phi_n^{a,b,c,d-1}(\by) \r: 
                 \nonumber \\  
     && \quad
          + a(a-1) \fr{3b_1}{4\pi^4} v_j (\bx-\by) \Phi_n^{a-2,b,c,d}(\by)
          + a b \fr{2}{\pi^2} u_j(\bx-\by) \Phi_n^{a-1,b-1,c,d}(\by)
                 \nonumber \\
     && \quad
          + b(b-1) \fr{12}{b_1} f_j(\bx-\by) \Phi_n^{a,b-2,c,d}(\by)
          + c(2c-1) \fr{2}{b_1} g_j(\bx-\by) \Phi_n^{a,b,c-1,d}(\by)
                 \nonumber \\
     && \quad
          + d \fr{4}{b_1}  \left\{ g_j(\bx-\by) - {\tilde g}_j(\bx-\by) \right\} \Phi_n^{a,b,c,d-1}(\by)
                 \nonumber \\
     && \quad
          + d(d-1) \fr{8}{b_1} e_{jk,l}(\bx-\by) \l: \pd^k \phi \pd^l \phi \Phi_n^{a,b,c,d-2}(\by) \r:
                 \nonumber \\
     && \quad
          - bd \fr{4}{b_1} f_{jk} (\bx-\by) \l: \pd^k \phi \Phi_n^{a,b-1,c,d-1}(\by) \r:
          - a n \fr{1}{4\pi^2} g_j(\bx-\by) \Phi_{n-1}^{a-1,b,c,d}(\by) 
                 \nonumber \\
     && \quad
          + b n \fr{1}{b_1} \left\{ g_j(\bx-\by) - {\tilde {\tilde g}}_j(\bx-\by) \right\} 
                              \Phi_{n-1}^{a,b-1,c,d}(\by) 
                 \nonumber \\
     && \quad
          + d n \fr{1}{b_1} e_{jk}(\bx-\by) \l: \pd^k \phi \Phi_{n-1}^{a,b,c,d-1}(\by) \r:
          - n(n-1) \fr{1}{2b_1} e_j(\bx-\by)  \Phi_{n-2}^{a,b,c,d}(\by) .
                 \nonumber \\
     &&
\end{eqnarray}
The other commutation relations are given by
\begin{eqnarray}
     && i \left[ \l: \P_\chi \pd_j \phi(\bx) \r:, \Phi_n^{a,b,c,d}(\by) \right] 
                 \nonumber \\
     && = 2c \dl_3 (\bx-\by) \l: \chi \pd_j \phi \Phi_n^{a, b, c-1,d}(\by) \r:
         + a c \fr{1}{2\pi^2} g_j(\bx-\by) \l: \chi \Phi_n^{a-1,b,c-1,d}(\by) \r:
                 \nonumber \\
     && \quad
         + b c \fr{2}{b_1}  g_j(\bx-\by) \l: \chi \Phi_n^{a,b-1,c-1,d}(\by) \r:
         + c d \fr{2}{b_1} g_{jk}(\bx-\by) \l: \chi \pd^k \phi \Phi_n^{a,b,c-1,d-1}(\by) \r:
                 \nonumber \\
     && \quad
         - c n \fr{1}{b_1} e_j(\bx-\by) \l: \chi \Phi_{n-1}^{a,b,c-1,d}(\by) \r: ,
                 \nonumber \\
     && i \left[ \l: \P_\chi \chi(\bx) \r:, \Phi_n^{a,b,c,d}(\by) \right]
                 \nonumber \\
     && = - (a - 2c) \dl_3 (\bx-\by) \l: \Phi_n^{a,b,c,d}(\by) \r: 
          - a(a-1) \fr{3b_1}{16\pi^4} u(\bx-\by) \Phi_n^{a-2,b,c,d}(\by)
                 \nonumber \\
     && \quad
           + a b \fr{1}{4\pi^2} u(\bx-\by) \Phi_n^{a-1,b-1,c,d}(\by) 
           + a d \fr{1}{2\pi^2} g_k(\bx-\by) \l: \pd^k \phi \Phi_n^{a-1,b,c,d-1}(\by) \:
                 \nonumber \\
     && \quad
           + c (2c-1) \fr{6}{b_1} g(\bx-\by) \Phi_n^{a,b,c-1,d}(\by)
           - a n \fr{3}{4\pi^2} g(\bx-\by)  \Phi_{n-1}^{a-1,b,c,d}(\by) ,
                 \nonumber \\
     && i \left[ \l: \chi^2(\bx) \r:,  \Phi_n^{a,b,c,d}(\by) \right] 
                 \nonumber \\
     && = - 2a \dl_3 (\bx-\by) \l: \chi \Phi_n^{a-1,b,c,d}(\by) \r: 
          - a c \fr{12}{b_1} g(\bx-\by) \l: \chi \Phi_n^{a-1,b,c-1,d}(\by) \r: ,
                 \nonumber \\
     && i \left[ \l: \chi \pd_j \phi(\bx) \r:,  \Phi_n^{a,b,c,d}(\by) \right] 
                 \nonumber \\
     && = - a \dl_3 (\bx-\by) \l: \pd_j \phi \Phi_n^{a-1, b, c,d}(\by) \r:
          - a(a-1) \fr{1}{4\pi^2} g_j(\bx-\by)  \Phi_n^{a-2,b,c,d}(\by)
                 \nonumber \\
     && \quad
          - a b \fr{1}{b_1}  g_j(\bx-\by) \Phi_n^{a-1,b-1,c,d}(\by) 
          - a d \fr{1}{b_1} g_{jk}(\bx-\by) \l: \pd^k \phi \Phi_n^{a-1,b,c,d-1}(\by) \r:
                 \nonumber \\
     && \quad
          + a n \fr{1}{2b_1} e_j(\bx-\by) \Phi_{n-1}^{a-1,b,c,d}(\by)
\end{eqnarray}
and also $[ \l: \pd_k \phi \pd^k \phi(\bx) \r:, \Phi_n^{a,b,c,d}(\by) ]=0$. The quantum correction terms are represented by the following functions: 
\begin{eqnarray}
    f(\bx) &=& -\fr{1}{40\pi^2}\fr{\eps(5\bx^2 -3\eps^2)}{(\bx^2 +\eps^2)^5} ,
            \qquad
    g(\bx) = -\fr{1}{6\pi^2} \fr{\eps}{(\bx^2 +\eps^2)^3} , 
            \nonumber \\      
    u(\bx) &=& \fr{1}{\pi^2} \fr{\eps(\bx^2 - 3 \eps^2)}{(\bx^2 +\eps^2)^5} , 
            \qquad
    f_j (\bx) = \fr{1}{\pi^2} \fr{\eps x_j (\bx^2 -\eps^2)}{(\bx^2 +\eps^2)^6} ,  
            \nonumber \\   
    g_j (\bx) &=& \fr{1}{\pi^2} \fr{\eps x_j}{(\bx^2 +\eps^2)^4} ,  
             \qquad
    {\tilde g}_j(\bx) = \fr{2}{\pi^2} \fr{\eps x_j[1 -h(\bx)]}{\bx^2 (\bx^2 +\eps^2)^3} ,
            \nonumber \\
    {\tilde {\tilde g}}_j(\bx) &=& \fr{6}{\pi^2} \fr{\eps x_j (\bx^2 -\eps^2)[1-h(\bx)]}{\bx^2 (\bx^2 +\eps^2)^4} ,
             \qquad
    e_j(\bx) = \fr{1}{\pi^2} \fr{\eps x_j [1-h(\bx)]}{\bx^2 (\bx^2 + \eps^2)^2} ,
             \nonumber \\
    u_j (\bx) &=& \fr{1}{\pi^2} \fr{\eps x_j \bx^2}{(\bx^2 +\eps^2)^6} ,  
             \qquad
    v_j(\bx) = \fr{1}{\pi^2} \fr{\eps x_j(\bx^2 - 3 \eps^2)]}{(\bx^2 +\eps^2)^6} ,
            \nonumber \\
    f_{jk} (\bx) &=&  - \fr{1}{\pi^2} \left\{ 
               \left( \dl_{jk} - \fr{3x_j x_k}{\bx^2} \right) 
                \fr{3 \eps (\bx^2 - \eps^2)[1-h(\bx)]}{\bx^2 (\bx^2 +\eps^2)^4}
               + \fr{x_j x_k}{\bx^2}\fr{\eps (5\bx^2 - 3 \eps^2)}{(\bx^2 +\eps^2)^5}  \right\} ,
            \nonumber \\
    g_{jk}(\bx) &=& \fr{1}{\pi^2} \left\{ \left( \dl_{jk} - \fr{3 x_j x_k}{\bx^2} \right) 
                          \fr{\eps [1-h(\bx)]}{\bx^2 (\bx^2 + \eps^2)^2} 
                          + \fr{x_j x_k}{\bx^2}\fr{\eps}{(\bx^2 +\eps^2)^3}  \right\} ,
             \nonumber \\
    e_{jk} (\bx) &=& \fr{1}{\pi^2} \left\{ 
             \left( \dl_{jk} - \fr{3x_j x_k}{\bx^2} \right) \fr{\eps [1-h(\bx)]}{\bx^2 (\bx^2 +\eps^2)^2}
             - \fr{x_j x_k}{\bx^2}\fr{\eps [3-4h(\bx)]}{(\bx^2 +\eps^2)^3}  \right\} ,
             \nonumber \\
    e_{jk,l} (\bx) &=& \fr{x_l}{\pi^2} \left\{ 
             \left( \dl_{jk} - \fr{3x_j x_k}{\bx^2} \right) \fr{\eps [1-h(\bx)]}{\bx^2 (\bx^2 +\eps^2)^3}
             + \fr{x_j x_k}{\bx^2}\fr{\eps}{(\bx^2 +\eps^2)^4}  \right\} ,
\end{eqnarray}
where $f_j(\bx) = \pd_j f(\bx)$ and $g_j(\bx) = \pd_j g(\bx)$ and the $h$ function is defined by
\begin{equation}
     h(\bx) = \fr{i\eps}{2|\bx|}\log \fr{i\eps + |\bx|}{i\eps -|\bx|} ,
          \label{function e_j and h}
\end{equation}
which satisfies $h^\dag(\bx)=h(\bx)$ and $\lim_{\bx \to 0} h(\bx) = 1$.

Here, we summarize the integrals of above quantum correction functions. The integrals of scalar functions are given by
\begin{eqnarray}
  && \int \d3x f(\bx) = 0,  \qquad   \int \d3x \bx^2 f(\bx) = - \fr{1}{160\eps^2}, 
           \nonumber \\  
  && \int \d3x g(\bx) = - \fr{1}{24\eps^2} ,  \qquad  \int \d3x u(\bx) = - \fr{3}{16 \eps^4} . 
\end{eqnarray}
The integrals of odd functions under the change $\bx \to -\bx$ vanish, while the integrals of such functions multiplied by $x_i$ are given by
\begin{eqnarray}
   && \int \d3x x_i f_j(\bx) = 0, \quad   \int \d3x x_i e_j(\bx) = \fr{1}{6} \dl_{ij} , 
              \nonumber \\
   && \int \d3x x_i g_j(\bx) = \int \d3x x_i {\tilde g}_j(\bx) = \int \d3x x_i {\tilde {\tilde g}}_j(\bx)
                             =  \dl_{ij} \fr{1}{24\eps^2} ,
              \nonumber \\
   && \int \d3x x_i u_j (\bx) = \dl_{ij} \fr{1}{128 \eps^4},   \qquad    
      \int \d3x x_i v_j (\bx) = - \dl_{ij} \fr{1}{64 \eps^4}.  
\end{eqnarray}   
Thus, we obtain the formulae $\int \d3x x_i \{ g_j(\bx) -{\tilde g}_j(\bx) \} = \int \d3x x_i \{ g_j(\bx) -{\tilde {\tilde g}}_j(\bx) \} = 0$. The other functions satisfy the following integral formulae:
\begin{eqnarray}
     && \int \d3x f_{jk} (\bx) = 0, \quad
        \int \d3x \bx^2 f_{jk} (\bx) = -\dl_{jk} \fr{1}{12\eps^2}, 
             \nonumber \\
     && \int \d3x x_j x^l f_{lk} (\bx) = -\dl_{jk} \fr{1}{24\eps^2}, \quad
        \int \d3x g_{jk} (\bx) = \dl_{jk} \fr{1}{12\eps^2}, \quad
             \nonumber \\
     && \int \d3x e_{jk}(\bx) = 0 ,  
        \int \d3x \bx^2 e_{jk}(\bx) =  -\fr{1}{3} \dl_{jk} ,  \quad 
        \int \d3x x_j x^l e_{lk}(\bx) = -\fr{2}{3} \dl_{jk} , 
            \nonumber \\
     && \int \d3x x_i e_{jk,l}(\bx) = \dl_{il} \dl_{jk} \fr{1}{60 \eps^2} 
            - \left( \dl_{ij} \dl_{kl} + \dl_{ik} \dl_{jl} \right) \fr{1}{240 \eps^2}  .
\end{eqnarray}

All quantum correction terms that diverge at $\eps \to 0$ cancel out in the computations given in the text. To show it, the following equations are useful:
\begin{eqnarray}
    \int \d3x \left\{ \bx^2 g_j(\bx-\by) - 2x_j x^k g_k(\bx-\by) -6x_j g(\bx-\by) \right\} &=& 0 ,
            \nonumber \\
    \int \d3x \left\{ \bx^2 e_j (\bx-\by) - 2x_j x^k e_k(\bx -\by) \right\} &=& -y_j ,
           \nonumber \\
    \int \d3x \left\{ \bx^2 e_{jk} (\bx-\by) - 2x_j x^l e_{lk}(\bx -\by) \right\} &=& \dl_{jk} ,
           \nonumber \\
    \int \d3x \left\{ \bx^2 f_{jk} (\bx-\by) - 2x_j x^l f_{lk}(\bx -\by) \right\} &=& 0 .
\end{eqnarray}


\end{document}